\documentclass[letterpaper, 10 pt, journal, twoside]{IEEEtran}

\usepackage{url}
\usepackage{amsmath} 
\usepackage{amssymb}  
\usepackage{amsthm}
\usepackage{lipsum,adjustbox}
\usepackage{tikz}
\usepackage[font=footnotesize]{caption}
\usepackage{multirow}

\usepackage[shortlabels]{enumitem}

\usepackage{cite}
\usepackage{graphicx,subcaption,float}

\DeclareMathOperator{\F}{\mathcal{F}}

\DeclareMathOperator{\K}{\mathcal{K}}
\DeclareMathOperator{\I}{\mathcal{I}}

\newcommand{\real}{\mathbb{R}}

\newcommand{\realnn}{\mathbb{R}_{\geq 0}}
\newcommand{\integer}{\mathbb{Z}}

\newcommand{\bz}{\bar{z}}

\newcommand{\zero}{\mathbf{0}}
\newcommand{\lie}{\mathcal{L}}
\newcommand{\until}[1]{\{1,\dots,#1\}}

\newcommand{\X}{\mathcal{X}}

\newtheorem{theorem}{Theorem}[section]
\newtheorem{proposition}[theorem]{Proposition}

\theoremstyle{remark}
\newtheorem{remark}{Remark}
\theoremstyle{definition}
\newtheorem{assumption}{Assumption}

\newcommand{\longthmtitle}[1]{\mbox{} \textit{(#1):}}
\newcommand{\setdef}[2]{\{#1 \; | \; #2\}}

\newcommand{\oprocendsymbol}{\hbox{$\square$}}
\newcommand{\oprocend}{\relax\ifmmode\else\unskip\hfill\fi\oprocendsymbol}

\graphicspath{{epsfiles/}}
\allowdisplaybreaks
\title{Agent-Supervisor Coordination for Decentralized Event-Triggered Optimization\thanks{This work was authored in part by the National Renewable Energy Laboratory (NREL), operated by Alliance
for Sustainable Energy, LLC, for the U.S. Department of Energy (DOE) under Contract DE-AC36-08GO28308 and supported by the Laboratory Directed Research and Development (LDRD) Program at NREL. The views expressed in the paper do not necessarily represent the views of the DOE or the U.S. Government. The U.S. Government retains and the publisher, by accepting the paper for publication, acknowledges that the U.S. Government retains a
nonexclusive, paid-up, irrevocable, worldwide license to publish or reproduce the published form of this work, or allow others to do so, for U.S. Government purposes.

This work was partially supported by NSF under Grant ECCS-1947050.
}}

\author{Priyank Srivastava \quad Guido Cavraro \quad Jorge Cort\'es
\thanks{P. Srivastava is with the Department of Mechanical Engineering, Massachusetts Institute of Technology. \texttt{psrivast@mit.edu}. 
G. Cavraro is with the National Renewable Energy Laboratory. \texttt{guido.cavraro@nrel.gov}.
J. Cort\'es is with the Department of Mechanical and Aerospace Engineering, University of California, San Diego. \texttt{cortes@ucsd.edu}.}
}

\begin{document}
\maketitle
\begin{abstract}
This paper proposes decentralized resource-aware coordination schemes for solving network optimization problems defined by objective functions that combine locally evaluable costs with network-wide coupling components.
These methods are well suited for a group of supervised agents trying to solve an optimization problem under mild coordination requirements. 
Each agent has information on its local cost
and coordinates with the network supervisor for information about the coupling term of the cost.  The proposed approach is feedback-based and asynchronous by design, guarantees anytime feasibility, and ensures the asymptotic convergence of the network state to the
desired optimizer. Numerical simulations on a power system example 
illustrate our results.
\end{abstract}

\thispagestyle{empty}


\section{Introduction}\label{sec:intro}
\IEEEPARstart{R}{ecent} advancements in digital systems, communication, and sensing technologies have enabled the deployment of large-scale cyber-physical systems in multiple domains.
Owing to the intrinsic modular structure of these systems, their applications can often be modeled as optimization problems over networks.
In many scenarios, the associated objective functions result from the combination of locally separable costs with global coupling terms whose evaluation requires members of a network layer to interact with an entity at a higher layer. From this arises the need for algorithmic implementations that help individual agents efficiently decide when to resort to such interactions. Motivated by these observations, this paper synthesizes resource-aware agent-supervisor coordination schemes that achieve asymptotic convergence to the desired network optimizer.

\emph{Literature Review:}
The present paper relies on two areas with significant recent activity: distributed optimization and event-triggered control. Because of the impossibility of surveying their vast literature, we provide the introductory references~\cite{TY:19-sv,PW-MDL:09,AN:14-sv} for the first  and~\cite{PT:07,WPMHH-KHJ-PT:12,LH-CF-HO-AS-EF-JPR-SIN:17} for the second.
A majority of work in network optimization builds on consensus-based approaches to solve problems 
where the overall objective is the summation of individual agents' private cost functions.
In such scenarios, each agent maintains and updates an estimate of the complete solution vector,
using local computation and peer-to-peer information exchange. An alternative architecture
has agents send and receive information from a central entity or a network supervisor. Such architecture
is well suited for scenarios where the objective function of the optimization problem is the combination of a component separable among the agents and another one coupling all the agents' decision variables. 
Notable examples of applications exhibiting this structure include virtual power plants~\cite{HS-MM-RT:11,GC-AB-RC-SZ:20}, where a central entity works as an aggregator enabling the participation of a cluster of distributed energy resources in the energy market; HVAC systems in intelligent buildings~\cite{SK-FB:13,JTW-SM:18}, where a central processor manages the main air supply and helps coordinate a group of thermal zones equipped with thermostats and controllers regulating heated or cooled air input; multi-agent systems with access to the cloud~\cite{MTH-AN-ME:18}, which provides superior  processing and storing capabilities; and sensor-actuator networks with a central computation node~\cite{MMJ-PT:10,HZ-PS-MJ:09} responsible for computing the control signal with measurements from distributed sensors. 
In these scenarios, there is a need to structure the interaction between individual agents and the network supervisor to design solutions that scale up.
Event-triggered control, see, e.g.,~\cite{PT:07,WPMHH-KHJ-PT:12,LH-CF-HO-AS-EF-JPR-SIN:17} and references therein, offers a framework to prescribe, in
a principled way, when to efficiently use the available resources while 
still guaranteeing a desired quality of service in performing the intended task. 
Several works~\cite{SSK-JC-SM:15-auto,CN-EG-JC:19-auto} have explored the use of event-triggered approaches for achieving network coordination tasks. Of particular relevance to our work are~\cite{MMJ-PT:10,SLB-CN-GJP:16,AA-DL-DVD-KHJ:15}, which consider similar network architectures, and~\cite{PO-JC:21-auto}, which mixes continuous updates, computed with the locally available information, with aperiodic updates, computed with the externally provided information.
An issue 
in event-triggered control
is the emergence of an infinite number of triggering times in a finite time interval, a.k.a. \emph{Zeno behavior}. 
Zeno behavior is addressed in~\cite{MMJ-PT:10,SLB-CN-GJP:16} by using \emph{time regularization}, i.e., preventing by design any update before a certain dwell time has elapsed. In general, time regularization requires an offline computation with global information. A final body of work we build on is that of continuous projected dynamical systems for optimization, cf.~\cite{TLF-DHB-NJM-RLT-SG:94,YSX-JW:00,XBG:03}.

\emph{Statement of Contributions:}
We consider network optimization problems where the objective function is the sum of a component given by the summation of local costs and a coupling component whose evaluation requires knowledge of all the agents' decision variables. Individual agents rely on the network supervisor to obtain information about the coupling component and to solve the optimization problem. 
Our contributions are structured in two blocks, corresponding to unconstrained and constrained problems. For unconstrained systems, we build on the gradient descent dynamics; for constrained systems with separable constraints, we build on globally projected dynamical systems. For both cases, we design novel event-triggered agent-supervisor coordination algorithms where agents continuously employ their local information and resort to opportunistic interactions with the supervisor for information about the coupling cost. The criterion for triggering employed by each agent depends only on locally available information, which makes the proposed approach suitable for applications where, due to privacy concerns, it is preferable not to share the local cost functions. We show the monotonic decrease of the objective function and establish the existence of a minimum inter-event time, thus ruling out Zeno behavior (without the need for any time regularization) and ensuring asymptotic convergence to the optimizer. In the constrained case, we also show that the feasible set is positively invariant, thus guaranteeing anytime feasibility. We illustrate the performance of the proposed algorithms with numerical simulations of a power system application.

\section{Preliminaries}\label{sec:prelims}
Here\footnote{
Throughout the paper, we employ the following notation.
Let $\real, \realnn$, and $\integer$ represent the set of real, nonnegative real, and integer numbers, respectively. 
$|\cdot|$ and $\|\cdot \|$ denote the absolute value of a scalar and the 2-norm of a vector, respectively.
$(\cdot)^\top$ denotes the transpose of a vector or a matrix.
A vector or a matrix of appropriate dimensions of all zero entries is denoted by $\zero$.
$A \succ \zero$ means that the matrix $A$ is positive definite.
A continuous function $\alpha:\realnn \to \realnn$ is of class
$\K_{\infty}$ if it is strictly increasing, $\alpha(0)=0$, and $\alpha(x)\to \infty$
as $x \to \infty$.
For a convex set $\Omega$, $\Pi_\Omega(y)$ denotes the projection of a point $y \in \real^n$ on~$\Omega$, i.e.,
$\Pi_\Omega(y)= \text{argmin}_{x \in \Omega} \|x-y\|$.}
we review the basic concepts on convex analysis, 
event-triggered control, and constrained optimization.

\subsubsection*{Event-Triggered Control}
The basics of event-triggered control following~\cite{PT:07,WPMHH-KHJ-PT:12} are presented next. 
Consider
\begin{align}\label{eq:nonlinear}
    \dot{x}=f(x,u),
\end{align}
where $x \in \real^n$ and $u \in \real^p$ denote the system state and input, respectively.
Assume there exists a control 
$u=k(x)$,
such that the 
closed-loop dynamics, 
\begin{align}\label{eq:closed_loop}
    \dot{x}=f(x,k(x+e)),
\end{align}
abbreviated $\psi_{\text{cl}}$, is input-to-state stable (ISS) with respect to the error  
$e \in \real^n$. Formally,
assume there exists a Lyapunov function $V$ such that its Lie derivative along~\eqref{eq:closed_loop} satisfies
\begin{align*}
    \lie_{\psi_{\text{cl}}}V \le -\alpha (\|x\|) + \gamma (\|e\|),
\end{align*}
where $\alpha$ and $\gamma$ are class $\K_{\infty}$ functions.
The implementation of the closed-loop system~\eqref{eq:closed_loop}
requires continuous updates of the actuator, which is not realizable in practice. Instead, event-triggered control seeks to prescribe opportunistic updates of the actuator that ensure the convergence properties of the closed-loop system are retained. This leads to a sample-and-hold implementation of the controller of the form
\begin{align}\label{eq:control_event}
u(t)=k(x(t_k)) \quad t \in [t_k, t_{k+1}),
\end{align}
where $\{t_k\}_{k=0}^{\infty}$ are the \emph{triggering times} when the control input is updated. To ensure the stability of the nonlinear system under~\eqref{eq:control_event} and to prescribe the triggering times, we look at the evolution of the Lyapunov function~$V$. Define the error variable as $e=x-x^k$, where we use the shorthand notation $x^k=x(t_k)$. 
With $\sigma \in (0,1)$, if the error satisfies
$\gamma (\|e\|) \le \sigma \alpha (\|x\|),$
then during the time $[t_k,t_{k+1})$,
$\lie_{\psi_{\text{cl}}} V \le (\sigma-1) \alpha (\|x\|).$
Note that at $t=t_k$, the error satisfies $e=0$; hence, setting $t_0=0$, we ensure $\lie_{\psi_{\text{cl}}} V < 0$ defining the triggering times as 
\begin{align}\label{eq:control_trigger}
    t_{k+1}=\min \setdef{t>t_k}{\gamma (\|e\|) = \sigma \alpha (\|x\|)}.
\end{align}
Although~\eqref{eq:control_trigger} guarantees the stability of the closed-loop system, Zeno behavior could arise.
Hence, for implementation in practice and to conclude asymptotic stability, 
it is necessary to have a uniform lower bound $\tau>0$ on the inter-event times, i.e., 
$t_{k+1}-t_k \ge \tau$ for all $k$. We refer to $\tau$ as the minimum inter-event time (\emph{MIET}). The existence of a MIET is guaranteed with the control law~\eqref{eq:control_event}
and the triggering condition~\eqref{eq:control_trigger} if the dynamics~\eqref{eq:nonlinear} is linear, and also for certain nonlinear systems under suitable assumptions, cf.~\cite{PT:07}.

\subsubsection*{Constrained Optimization via Continuous Projected Dynamical Systems}
Here we review the basic concepts on the stability of  continuous projected dynamical systems and their application in constrained optimization following~\cite{YSX-JW:00,XBG:03}.
Consider the following optimization problem
\begin{align}\label{eq:constrained}
    \min\limits_{x \in \Omega} \; h(x),
\end{align}
where $h: \real^n \to \real$ is a continuously differentiable function and $\Omega \subseteq \real^n$ is a convex set. 
Problem~\eqref{eq:constrained} can be solved using
\begin{align}\label{eq:projected}
    \dot{x}=\Pi_\Omega (x -  \lambda \nabla h(x)) - x,
\end{align}
where $\lambda > 0$ is a design parameter.
Unlike the commonly employed projected gradient dynamics, cf.~\cite{UH-JBM:94},
which is discontinuous at the boundary of $\Omega$, the dynamics~\eqref{eq:projected} is continuous due to the gradual application of the projection operator throughout the constraint set.
The following result characterizes its convergence properties.

\begin{theorem}\longthmtitle{Forward invariance of the feasible set and convergence to an optimizer~\cite{YSX-JW:00,XBG:03}}\label{thm:invariant}
Assume that $\nabla h$ is locally Lipschitz continuous on an open set containing $\Omega$. Then
\begin{enumerate}[(i)]
    \item The solution of~\eqref{eq:projected} approaches the set $\Omega$ exponentially fast. Moreover, if $x(0) \in \Omega$, then $x(t) \in \Omega$ for all $t > 0$.
    \item For all $\lambda > 0$, the dynamics~\eqref{eq:projected} is stable,
and for any initial condition $x(0) \in \Omega$, the trajectory of~\eqref{eq:projected} converges to a solution of~\eqref{eq:constrained}.
\end{enumerate}
\end{theorem}

\section{Problem Formulation}\label{sec:problem}
Consider a network of $n \in \integer$ agents and a  supervisor,
collectively seeking to solve 
\begin{align}\label{eq:problem}
\min_{x \in \X} \quad \underbrace{\sum\limits_{i=1}^n f_i(x_i)}_{f(x)} + \; g(x),
\end{align}
where, for all $i \in \until{n}$, $f_i: \real \to \real$ is the local cost
function of agent $i$, 
$g:\real^n \to \real$ is a function coupling all the agents' states, 
$\X=\prod\limits_{i=1}^n \X_i$ is the constraint set, and $\X_i$ is agent~$i$'s constraint set.
Each agent $i \in \until{n}$ has knowledge of its local state $x_i \in \real$, 
constraint set $\X_i$ and cost $f_i$, 
and relies on  the network supervisor to obtain information pertaining the coupling cost~$g$. 
For simplicity of exposition, we assume that~\eqref{eq:problem} has a unique solution $x^*$,
albeit the results of the paper can be extended with minor modifications to the case of multiple optimizers. We make the following assumptions on the cost functions and the constraints.

\begin{assumption}\longthmtitle{Convexity and Lipschitz gradients}
The functions $\{f_i\}_{i=1}^n$ and $g$ are convex; 
the functions $\{f_i\}_{i=1}^n$ are twice continuously differentiable, 
and $g$ is continuously differentiable with locally Lipschitz gradient; and the sets $\{\X_i\}_{i=1}^n$ are compact and convex.
\end{assumption}

Our goal is to design a decentralized algorithm that allows the agents to collectively solve~\eqref{eq:problem}. 
We want the algorithm to be anytime, meaning that if the network state starts feasible, it remains so during the algorithm's execution.  This anytime nature is desirable in applications where the optimization problem is not stand-alone and its solution serves as an input to another layer in the control design. 
In such cases, the algorithm should yield a feasible solution even if terminated in finite time.
Note that, without the presence of the coupling function~$g$,~\eqref{eq:problem} could be solved easily by having each agent $i$ solve a local optimization problem with the function $f_i$ over the constraint set~$\X_i$.
Instead, the presence of $g$ couples the agents' decisions. 
Since information about $g$ is not available at all times, we seek to endow the individual agents with a criterion that allows them to determine when to query the supervisor in an opportunistic fashion -- this is what corresponds to the event-triggered component of the algorithmic solution. 
Based on the application at hand and the supervisor's capabilities, the coordination between the supervisor and the agents could be
feedback-based or computation-based:

(i) \emph{Sensing-based:} each agent $i \in \until{n}$ can evaluate $\nabla_{x_i} g$ with its local information and the one broadcast from the supervisor,
when an agent asks for an update. 
This is because the supervisor has access from its own measurements to enough knowledge about~$g$. 
This is common in cyber-physical scenarios where access to field measurements of the physical layer provides global information about the network state (for instance, in power systems, see e.g.~\cite{MC-EDA-AB:20}, local measurements of the frequency deviation provide information about the overall network mismatch in meeting the prescribed load).
In this case, the states of the agents remain private (the virtual power plant in~\cite{GC-AB-RC-SZ:20} is an example falling in this case).

(ii) \emph{Computation-based:} the supervisor knows the functional form of the cost.
Whenever an agent asks for an update, the supervisor gathers the state of all the agents, evaluates $\nabla g$, and broadcasts it to the agents (scheduling links and channels for transmission of information in wireless networks is an example scenario for this case, see, e.g.,~\cite{HZ-PS-MJ:09}).

The forthcoming design and the ensuing analysis can be applied to both scenarios.
\begin{remark}\longthmtitle{Synchronicity of the updates}
The coordination between the agents and the supervisor described above
requires the broadcast of information to all the agents whenever any agent asks for an update.
As a result, $\{\nabla_{x_i} g(x)\}_{i=1}^n$ are updated synchronously.
From a practical viewpoint, this is reasonable in the sensing-based scenario because once the supervisor has the required information available, it can broadcast it to all the agents. Instead, in the computation-based scenario,  
the supervisor might not have up to date information about all the agents. In this case, we assume that the gathering of information by the supervisor can be done
simultaneously from all the agents whenever there is an update request. \hfill $\bullet$
\end{remark}

\section{Event-Triggered Coordination for Unconstrained Problems}\label{sec:unconstrained}
Here, an event-triggered decentralized algorithm to solve~\eqref{eq:problem} when $\X=\real^n$ is provided. 
Consider the standard gradient-descent dynamics
\begin{align*}
\dot{x}= - \lambda (\nabla f(x) + \nabla g(x)),
\end{align*}
where $\lambda>0$ is a design parameter. For agent $i \in \until{n}$, this takes the form
\begin{align}\label{eq:gradient_i}
\dot{x}_i= - \lambda (\nabla_{x_i} f_i(x_i) + \nabla_{x_i} g(x)) .
\end{align}
From an implementation viewpoint, the first term in~\eqref{eq:gradient_i} can be evaluated locally by each agent,
the second term, however,
entails 
continuous communication with the supervisor. 
We avoid this by designing an event-triggered scheme that has the supervisor broadcast the information needed to compute $\nabla g(x)$ in an opportunistic fashion.
With the shorthand notation $x^k=x(t_k)$, consider the dynamics
\begin{align}\label{eq:event}
\dot{x}= -\lambda (\nabla f(x) + \nabla g(x^k)) \quad t_k \le t < t_{k+1}.
\end{align}
To implement~\eqref{eq:event}, the network supervisor needs to broadcast the information required to compute $\nabla g(x)$ only at some specified times $\{t_k\}_{k=0}^\infty$.
Here, $\nabla g(x^k)$ is the equivalent of the input in the standard event-triggered control, cf.~Section~\ref{sec:prelims}.
When convenient, we refer to the dynamics~\eqref{eq:event} as ${\psi_\text{ev}}$.
The next result identifies a decentralized condition on the triggering times $\{t_k\}_{k=0}^\infty$ that ensures that the dynamics~\eqref{eq:event} is stable. By decentralized, we mean that each agent $i \in \until{n}$ can identify the triggering criterion locally without knowing the states of the other agents or the coupling function.

\begin{proposition}\longthmtitle{Decentralized trigger}\label{prop:trigger}
Let $x^k \neq x^*$ be the state when the trigger was last implemented,
$\sigma\in (0,1)$, and define $\F_k = \setdef{x \in \real^n}{f(x)+g(x) \le f(x^k)+g(x^k)}$.
Then for all $\lambda > 0$, the dynamics~\eqref{eq:event} is stable and the value of the objective function $f+g$ is nonincreasing if the triggering times are updated according to 
\begin{align}\label{eq:trigger}
t_{k+1}&=\min \limits_{i \in \until{n}} \min \setdef{t>t_k}{  
\notag
\\
 \qquad & L_g|x_i  - x^k_i| = \sigma |\nabla_{x_i} f_i(x_i) + \nabla_{x_i}g(x^k)| \ne 0} , 
\end{align}
where $L_g$ is the Lipschitz constant of $\nabla g$ over $\F_0$.
\end{proposition}
\begin{IEEEproof}
Consider the Lyapunov function $V:\real^n \to \real$
\begin{align}\label{eq:V}
    V(x)=&f(x)+g(x)-f(x^*)-g(x^*),
    \end{align}
    whose Lie derivative is
    \begin{align*}
    \lie_{\psi_\text{ev}} V (x) = -\lambda (z+ e)^\top  z  
     \le - \lambda \|z \|^2\left( 1 - \frac{\|e\|}{\|z\|} \right),
\end{align*}
where $z=[z_1 \; \ldots \; z_n]^\top$, $z_i = \nabla_{x_i} f_i(x_i) + \nabla_{x_i}g(x^k)$, and $e=\nabla g(x) - \nabla g(x^k)$.
At $t=t_k$, we have $e=0$; then the error starts increasing as $x^k$ becomes obsolete. However, $\lie_{\psi_\text{ev}} V \le 0$ if we ensure that $\|e\| \le \sigma \| z \|$. 
The direct evaluation of the latter condition requires complete information about the network state.
However, note that $
    \|e\| \le  L_g^k \|x-x^k\|$, where $L_g^k$ is the Lipschitz constant of $\nabla g$ over $\F_k$.
Hence, we can guarantee the stability of~\eqref{eq:event} if 
\begin{align}\label{eq:trigger_central}
     L_g^k\|x-x^k\| \le \sigma \| z \|.
\end{align}
The triggering rule~\eqref{eq:trigger} ensures that~\eqref{eq:trigger_central} is satisfied noting that the set $\F_k$ is forward invariant, $\F_{k+1} \subseteq \F_k$, and hence
$L_g \ge L_g^k$ for all $k$.
\end{IEEEproof}

From Proposition~\ref{prop:trigger}, it is clear that if the agents have knowledge of (an upper bound on) $L_g$,
they can check~\eqref{eq:trigger} locally and request the supervisor for an update accordingly.
Although~\eqref{eq:trigger} guarantees that the dynamics~\eqref{eq:event} is stable, 
we still need to establish the convergence to $x^*$ and whether the proposed event-triggered scheme is Zeno-free.  We prove both facts in the next result. 

\begin{proposition}\longthmtitle{Non-Zeno behavior and convergence to the optimizer}\label{prop:zeno}
With the notation of Proposition~\ref{prop:trigger}, 
if the triggering times are updated as~\eqref{eq:trigger}, then for all $\lambda>0$, the MIET is lower bounded by 
$\tau = \frac{1}{\lambda H} \log (\sigma \lambda H/L_g+1 ) > 0,$
where $H = \max\limits_{i \in \until{n}} \max\limits_{x \in \F_0} \nabla^2_{x_i} f_i(x_i)$.
Moreover, any trajectory of~\eqref{eq:event} converges asymptotically to $x^*$.
\end{proposition}
\begin{IEEEproof}
If $x^k=x^*$, then the result is immediate. Assume then that $x^k \ne x^*$.
Let $\I = \setdef{i}{z_i(t_k) \ne 0}$. 
Since for all $i \in \until{n}$, 
$ \dot{z}_i = - \lambda \nabla^2_{x_i} f_i(x_i) z_i$, we deduce that $z_i=0$ for all $t \in [t_k,t_{k+1})$ if $i \notin \I$. For $i \in \I$, we
examine the evolution of $|x_i - x^k_i|/|z_i|$,
\begin{align}\label{eq:trigger_ratio}
    \frac{d}{dt} \frac{|x_i - x^k_i|}{|z_i|}
    &= \frac{(x_i-x^k_i) z_i}{\sqrt{(x_i - x^k_{i})^2}\sqrt{z_i^2}} - \frac{z_i \dot{z}_i\sqrt{(x_i - x^k_{i})^2}}{|z_i|^3} \notag
        \\
     \le & \; 1 + \frac{|\dot{z}_i|}{|z_i|} \frac{|x_i - x^k_i|}{|z_i|}  \le 1 + \lambda H \frac{|x_i - x^k_i|}{|z_i|}.
\end{align}
Now consider the differential equation
$\dot{y}=1+ \lambda H y$, $t_k \le t < t_{k+1}$, with initial condition $y(t_k)=0$, whose closed-form solution is given by
\begin{align*}
    y=\frac{1}{\lambda H} \big(e^{\lambda H(t-t_k)} -1\big), \quad t_k \le t < t_{k+1}.
\end{align*}
By the Comparison Principle, cf.~\cite[Lemma 3.4]{HKK:02}, we have
\begin{align*}
    \frac{|x_i - x^k_i|}{|z_i|} \le \frac{1}{\lambda H} \big(e^{\lambda H(t-t_k)} -1\big) , \quad t_k \le t < t_{k+1}.
\end{align*}
Equating the right-hand side of the above inequality with $\sigma / L_g$ implies that the inter-event time is lower bounded by~$\tau$ 
provided $z_i \ne 0$ for all $t \in [t_k,t_{k+1})$ and each $i \in \I$.
We reason by contradiction to prove this.
Since the ratio $|x_i-x_i^k|/|z_i|$ is bounded,
$z_i = 0$ only if $x_i-x_i^k=0$.
Let $\bar{t} = \min \setdef{t>t_k}{x_i-x_i^k=0}$.
Since $x_i-x_i^k=0$ and $z_i \ne 0$ at $t=t_k$, this means that
the sign of $z_i$ has to change before $\bar{t}$, and from the continuity of the dynamics,
there exists $\hat{t} < \bar{t}$ such 
$z_i(\hat{t})=0$, which contradicts $z_i \ne 0$ for all $t \in [t_k, \bar{t})$.
To prove the attractivity part, note that from Proposition~\ref{prop:trigger}, $\lie_{\psi_\text{ev}} V \le 0$,
and hence, $\lie_{\psi_\text{ev}} V(x) < 0$ for all $x \ne x^*$ as $z_i \ne 0$ for all $i \in \I$
and all $t \in [t_k, t_{k+1})$.
\end{IEEEproof}

\begin{remark}\longthmtitle{Differentiability of the local objective functions}
Note that ruling out Zeno behavior in Proposition~\ref{prop:zeno} relies on 
the functions $\{f\}_{i=1}^n$ being twice continuously differentiable, whereas
the dynamics~\eqref{eq:event} and the triggering condition~\eqref{eq:trigger} 
involve only first-order derivatives. 
We believe, although we do not pursue it here for space reasons, that Proposition~\ref{prop:zeno} can be extended for the case when the separable component of the objective function is just continuously
differentiable, using tools from nonsmooth analysis, e.g.,~\cite{FHC:83,JC:08-csm}. \hfill $\bullet$
\end{remark}

\begin{remark}\longthmtitle{Self-triggered implementation}
In the absence of errors in the solution of the differential equations by the individual agents, the criterion~\eqref{eq:trigger} can also be implemented in a self-triggered fashion. In fact, we can write it as 
\begin{align*}
t^i_{k+1} &= 
\min \setdef{t>t_k}{ 
L_g|x_i  - x^k_i| =
\notag
\\ & \qquad
\sigma |\nabla_{x_i} f_i(x_i) + \nabla_{x_i}g(x^k)| \ne 0} ,
 \\
  t_{k+1}&=\min\limits_{i \in \until{n}} t_{k+1}^i.
\end{align*}
This means that, with the information provided at time $t_k$, 
each agent $i\in \until{n}$ can compute $t^i_{k+1}$ by solving its differential equation, 
and convey it to the supervisor, which can then schedule the next triggering event at~$t_{k+1}$.
\hfill $\bullet$
\end{remark}

\section{Event-Triggered Coordination for Constrained Problems}\label{sec:constrained}
To deal with constrained problems, we build on the continuous projected dynamics~\eqref{eq:projected}, which takes the form
\begin{align*}
    \dot{x}=\Pi_\X(x- \lambda(\nabla f(x)+\nabla g(x))) -x,
\end{align*}
where $\lambda > 0$.
Its event-triggered counterpart is
\begin{align}\label{eq:cont_event}
    \dot{x}=\Pi_\X(x-\lambda(\nabla f(x)+\nabla g(x^k))) -x,
\end{align}
for $t \in [t_k,t_{k+1})$.
When convenient, we refer to the dynamics~\eqref{eq:cont_event} as ${\psi_{\text{evco}}}$.
The following result identifies a decentralized condition on the triggering times $\{t_k\}_{k=0}^\infty$ that ensures the stability of~\eqref{eq:cont_event}. 

\begin{proposition}\longthmtitle{Decentralized trigger for constrained problems}\label{prop:trigger_constraints}
Let $x^k \neq x^*$ be the state when the trigger was last implemented, 
and $\sigma \in (0,1)$.
If $x(0) \in \X$, then for all $\lambda > 0$, $x(t) \in \X$ for all $t > 0$,
the dynamics~\eqref{eq:cont_event} is stable and the value of the objective function $f+g$ is non-increasing if
 the triggering times are updated as 
\begin{align}
t_{k+1}=&\min\limits_{i \in \until{n}}  \min \setdef{t>t_k}{  \lambda \bar{L}_g|x_i - x^k_i| = \notag 
\\ \sigma  | \Pi_{\X_i} & (x_i-\lambda(\nabla_{x_i} f_i(x_i)+\nabla_{x_i} g(x^k))) -x_i | \ne 0}, \label{eq:trigger_constraints}
\end{align}
where $\bar{L}_g$ is the Lipschitz constant of $\nabla g$ over $\X$.
\end{proposition}
\begin{IEEEproof}
We start by noting that from Theorem~\ref{thm:invariant},  
for $t \in [t_k,t_{k+1})$, 
positive invariance of the feasible set $\X$ under~\eqref{eq:cont_event}
can be established by taking $h(x) \equiv f(x)+\nabla g(x^k)^\top x$.
To prove stability, consider again the Lyapunov function candidate $V$ defined in~\eqref{eq:V},
whose Lie derivative is now given by
    \begin{align*}
    \lie_{\psi_{\text{evco}}} V (x)= & (\nabla f(x) + \nabla g(x^k)+e)^\top \bz,
    \end{align*}
    where $\bz=[\bz_1 \; \ldots \; \bz_n]^\top$, $\bz_i=\Pi_{\X_i}(x_i-\lambda(\nabla_{x_i} f_i(x_i)+\nabla_{x_i} g(x^k))) -x_i$, and $e=\nabla g(x) - \nabla g(x^k)$.
    It is well known, cf.~\cite{XBG:03}, that for a convex set $\Omega$
    \begin{align*}
        (u-\Pi_\Omega(u))^\top (\Pi_\Omega(u) - v) \ge 0,
    \end{align*}
    for all $v \in \Omega$ and all $u \in \real^n$. 
    With $\Omega=\X$, $v=x$, and $u=x-\lambda(\nabla f(x)+ \nabla g(x^k))$, this implies that
    \begin{equation*}
        (\lambda\nabla f(x)+ \lambda \nabla g(x^k) + \bz) ^\top \bz \le 0.
    \end{equation*}
Using this, we upper bound the Lie derivative as 
    \begin{equation*}
        \lie_{\psi_{\text{evco}}} V (x) \le -\frac{1}{\lambda} \bz^\top \bz + e^\top \bz 
        \le -\|\bz\|^2 \left( \frac{1}{\lambda} - \frac{\|e\|}{\|\bz\|} \right).
\end{equation*}
This expression is analogous to that of $\lie_{\psi_{\text{ev}}} V$ in the proof of Proposition~\ref{prop:trigger}, and a similar argument concludes the proof.
\end{IEEEproof}

As in the unconstrained case, 
without excluding Zeno behavior, Proposition~\ref{prop:trigger_constraints} is not enough to conclude the asymptotic convergence to $x^*$. 

\begin{proposition}\longthmtitle{Non-Zeno behavior and convergence to the optimizer in the constrained case}\label{prop:zeno_constraints}
With the notation of Proposition~\ref{prop:trigger_constraints}, 
if the triggering times are updated as~\eqref{eq:trigger_constraints}, 
then for all $\lambda < \bar{\lambda}= 1/\bar{H}$, the MIET is lower bounded by $\bar{\tau} = \log (\sigma/\lambda \bar{L}_g + 1) > 0$,
where $\bar{H} = \max\limits_{i \in \until{n}} \max\limits_{x_i \in \X_i} \nabla^2_{x_i} f_i(x_i)$.
Moreover, any trajectory of~\eqref{eq:cont_event} with $x(0) \in \X$ converges asymptotically to $x^*$.
\end{proposition}
\begin{IEEEproof}
Since $\{\X_i\}_{i=1}^n$ are compact and convex,
without loss of generality let $\X_i=\setdef{x_i \in \real}{\underline{x}_i \le x_i \le \overline{x}_i}$ for all $i$.
For each agent $i \in \until{n}$, define $u_i: \X_i \to \real$
as $u_i(x_i)=x_i - \lambda (\nabla_{x_i} f_i(x_i) + \nabla_{x_i} g(x^k))$. 
The derivative of $u_i$ w.r.t $x_i$ is given by
\begin{align*}
    \frac{d u_i}{d x_i}=1 - \lambda \nabla^2_{x_i} f_i(x_i).
\end{align*}
For a given $i \in \until{n}$, the sign of $du_i/dx_i$ at any $x_i \in \X_i$ depends on the value of $\lambda$ and $\nabla^2_{x_i} f_i(x_i)$.
If $\lambda<\bar{\lambda}$, then $du_i/dx_i > 0$ for all $x_i \in \X_i$.
This means that if there is a point $\hat{x}_i \in \X_i$ such that
$\Pi_{\X_i} (u_i(\hat{x}_i)) = \overline{x}_i $, then $\Pi_{\X_i} (u_i(x_i)) = \overline{x}_i $ for all $x_i > \hat{x}_i$.
Similarly, if there is a point $\tilde{x}_i \in \X_i$ such that
$\Pi_{\X_i} (u_i(\tilde{x}_i)) = \underline{x}_i $, then $\Pi_{\X_i} (u_i(x_i)) = \underline{x}_i $ for all $x_i < \tilde{x}_i$.
Therefore, $\dot{\bz}_i$ can be represented as a set-valued map, cf.~\cite{JC:08-csm},
\begin{align*}
    |\dot{\bz}_i|=\begin{cases} |\lambda \nabla^2_{x_i} f_i(x_i) \bz_i| & \quad \tilde{x}_i < x_i < \hat{x}_i, \\ 
    [| \lambda \nabla^2_{x_i} f_i(x_i) \bz_i|,|\bz_i|] & \quad x_i=\tilde{x}_i,\hat{x}_i, \\
    |\bz_i| & \quad  \underline{x}_i \le x_i < \tilde{x}_i, \;  \hat{x}_i < x_i \le \overline{x}_i. \end{cases}
\end{align*}
Since $\lambda < \bar{\lambda}$, we have an expression similar to~\eqref{eq:trigger_ratio} for $\frac{d}{dt}|x_i-x_i^k|/|\bz_i|$ 
for all $i$ with $\bz_i(t_k) \ne 0$, with $\lambda H$ replaced by 1. The remainder of the argument follows analogously to the proof of Proposition~\ref{prop:zeno}.
\end{IEEEproof}
The upper bound on $\lambda$ in Proposition~\ref{prop:zeno_constraints}
is conservative in general.
In fact, the dynamics~\eqref{eq:cont_event} with the triggering rule~\eqref{eq:trigger_constraints} 
may be Zeno-free even if this condition is not satisfied, something that we have observed in simulation.

 \section{Simulations}
 
 \begin{figure*}[htb]
   \begin{subfigure}[htb]{.25\linewidth}{\includegraphics[width=\linewidth]{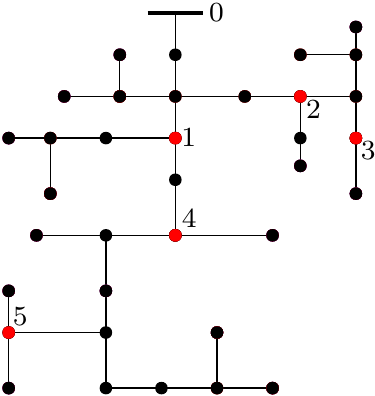}}
  \caption{}
  \end{subfigure}
 \begin{subfigure}[htb]{.375\linewidth}{\includegraphics[width=\linewidth]{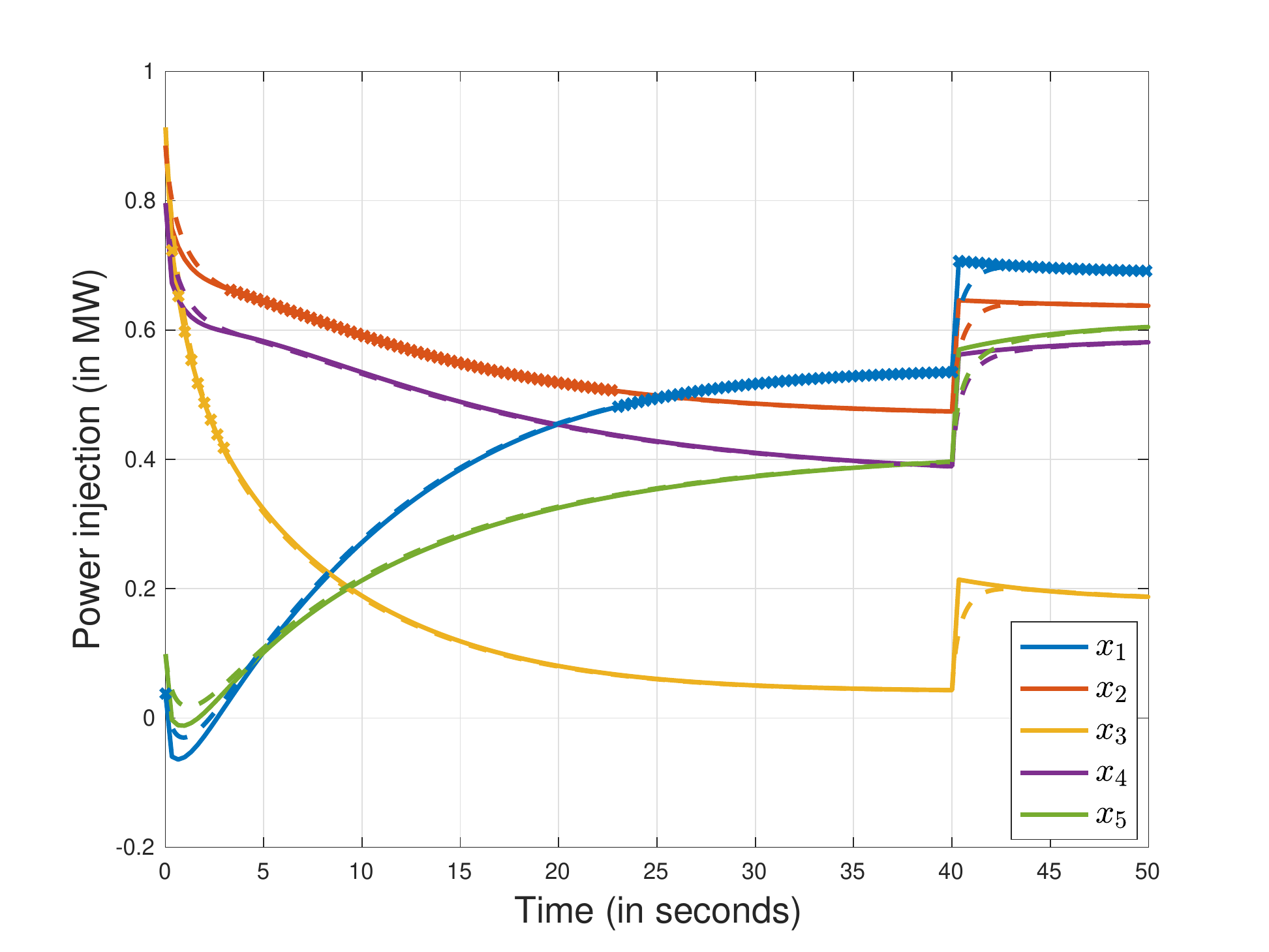}}
  \caption{}
  \end{subfigure}\hspace{-1ex}
  \begin{subfigure}[htb]{.375\linewidth}{\includegraphics[width=\linewidth]{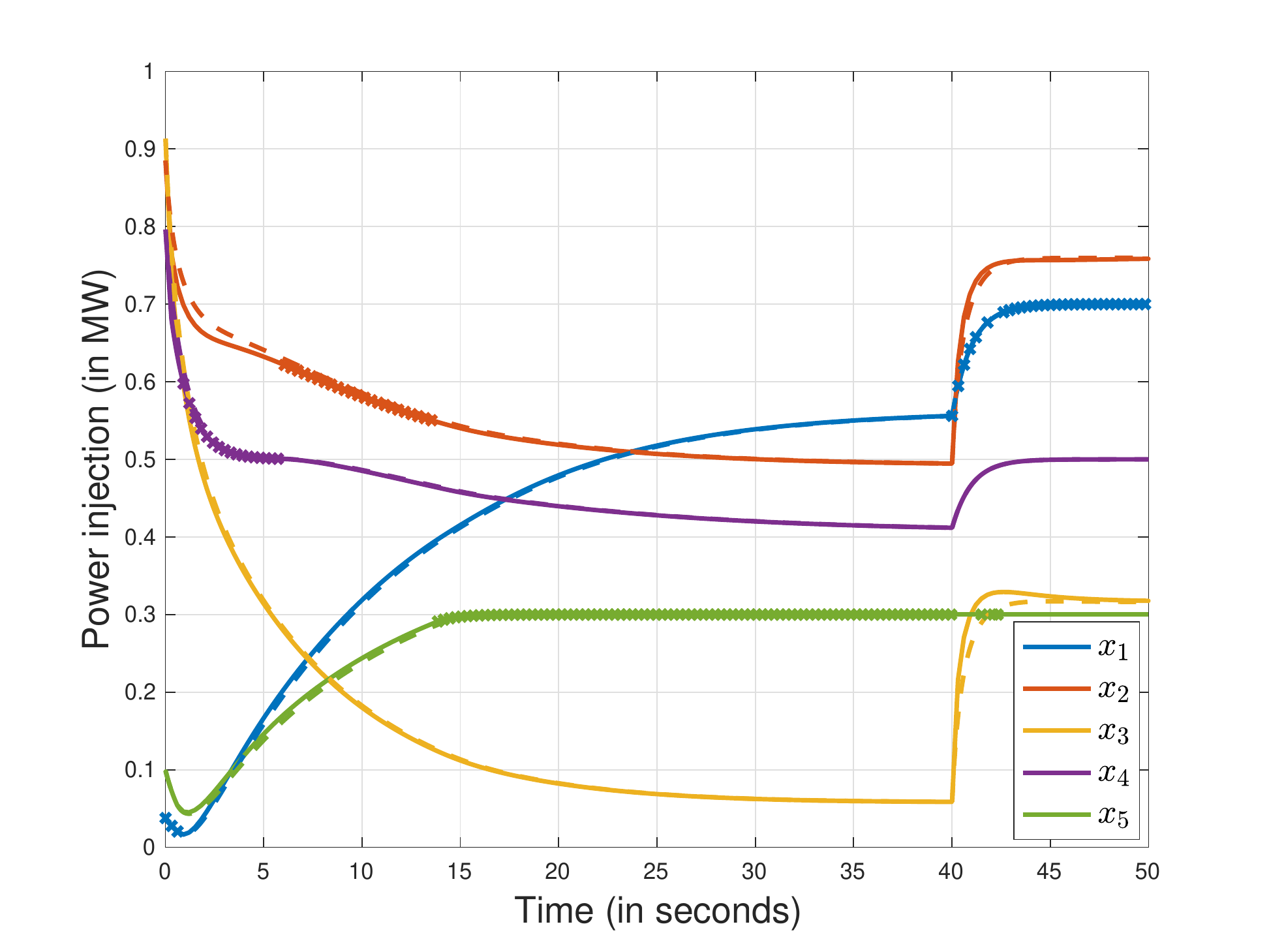}}
  \caption{}
  \end{subfigure}
  \caption{Performance of proposed event-triggered coordination algorithms on the power distribution network.  (a) shows the IEEE 37-bus test feeder, where node 0 represents the supervisor,
  red nodes (1-5) represent the generators, and black nodes represent the loads; edges represent the electrical connection between the nodes.   The total load increases by 1MW after 40 seconds.
  The values of $\lambda$ and $\gamma$ are taken as 0.2 and 0.9, respectively and stepsize is taken as $10^{-2}$. 
  (b) and (c) compare the state evolution using the proposed event-triggered mechanisms (solid lines) 
  using the triggering rules~\eqref{eq:trigger} and~\eqref{eq:trigger_constraints} with their respective continuous update cases.
  $\times$ markers denote the triggering instances for the corresponding agent.}
  \label{fig:sims}
  \end{figure*}

Here we test our coordination approach in a power distribution scenario, where $n$ generators managed by a distribution system operator that acts as the network supervisor seek to minimize the total power generation cost.
The active power output of generator $i=\until{n}$ is~$x_i$ and
the power that flows from the external grid to the distribution grid through the substation, labeled as node 0, is denoted as~$x_0$.
Assuming the power line losses are negligible, $x_0$ can be linearized as 
\begin{equation}
    x_0(x) \simeq - \sum_{i=1}^n x_i + c,
    \label{eq:p0_appx}
\end{equation}
where $c$ depends on the grid load, see, e.g.,~\cite{GC-AB-RC-SZ:20}.
We choose polynomial costs for power injection
at each node $i = \{0,1,\dots n\}$.
Each function $f_i$ is known only by generator $i \in \until{n}$, and the network supervisor has access only to $f_0$.
By substituting~\eqref{eq:p0_appx} into $f_0(x_0)$, the problem of minimizing the total power cost 
is equivalent to solving~\eqref{eq:problem} with $g(x)=f_0(- \sum_{i=1}^n x_i + c)$.
From the chain rule, $\nabla_{x_i}g(x) = -\nabla_{x_0}f_0(x_0)$,
for all $i \in \until{n}$.
Since the supervisor can measure $x_0$,
whenever there is an update request,
it can evaluate $\nabla g$ directly and broadcast it to the generators. This corresponds to the feedback-based scenario (cf. Section~\ref{sec:problem}).

We test the algorithms resulting from the triggering criteria~\eqref{eq:trigger} and~\eqref{eq:trigger_constraints} on a single-phase equivalent of the IEEE 37-bus test feeder, reported in Figure~\ref{fig:sims}(a).
The network has five generators. 
The load buses are a mixture of constant-current, constant-impedance, and constant-power loads~\cite{WHK:01}.
The initial active and reactive power demands are 2~MW and 1~MVAR, respectively.
At $t=40$ seconds, the active power demand increases by 1~MW.
The algorithms are simulated using the nonlinear exact AC power flow solver MATPOWER~\cite{RDZ-CEM-DG:11}.

We provide two sets of simulations based on whether the generation capacities of the generators are constrained or unconstrained. 
For the constrained case, for each $i \in \until{5}$, $\X_i=\setdef{x_i}{0 \le x_i \le \overline{x}_i}$,
where $\{\overline{x}_i\}_{i=1}^5$ are taken as 0.7~MW, 1~MW, 0.8~MW, 0.5~MW and 0.3~MW, respectively.
Figure~\ref{fig:sims}(b)-(c) show the evolution of the injected active powers of the agents using the proposed event-triggered dynamics
and the continuous dynamics
for the unconstrained and constrained case, respectively.
Note that depending on the operating region, different agents request for the information update non-uniformly.
For a fixed load, we simulate 10 different initial conditions for $60$ seconds (sec) and
observe an average MIET of
$3.3 \times 10^{-1}$ sec for the unconstrained and $2.0 \times 10^{-1}$ sec for the constrained case,
with respective standard deviations of $6.7 \times 10^{-3}$ sec and $23.1 \times 10^{-3}$ sec, and 
$177$ (unconstrained) and $190$ (constrained) average updates per resource-aware execution. 

\section{Conclusions and Future Work}
We have designed decentralized event-triggered coordination mechanisms to solve network optimization problems whose objective function is a combination of a separable component among the agents and a non-separable coupling term. The proposed coordination mechanisms 
prescribe opportunistic requests for information from the agents to the network supervisor, are anytime, and guarantee asymptotic convergence to the desired optimizer. 
Future work will focus on synthesizing novel 
adaptive 
triggering criteria ensuring uniform participation of all the agents and minimizing the 
total number of updates, 
consider more general interaction topologies, optimization problems with locally-coupled constraints, 
and asynchronous updates in computation-based scenarios, where the network supervisor receives only fresh state information from the agent triggering the update and as a result evaluates the information about the coupling term with outdated knowledge of the overall network state.

\end{document}